\documentclass[12pt,oneside]{article}
\pdfoutput=1
\newenvironment{wileykeywords}{\textsf{Keywords:}\hspace{\stretch{1}}}{\hspace{\stretch{1}}\rule{1ex}{1ex}}

\usepackage{amsmath,amssymb}
\usepackage{graphicx}
\usepackage{color}
\usepackage{dcolumn}
\usepackage{bm}
\usepackage{cancel}
\usepackage{soul}
\usepackage[normalem]{ulem}
\usepackage{bigints}

\usepackage[numbers,super,comma,sort&compress]{natbib}

\definecolor{background-color}{gray}{0.98}

\usepackage[raggedright,loose]{subfigure}

\newcommand{\fig}{Fig.}

\newcommand{\figref}[1]{\fig~\ref{#1}}

\renewcommand{\eqref}[1]{Equation~(\ref{#1})}
\newcommand{\Eqref}[1]{Equation~(\ref{#1})}

\newcommand{\bfy}{{\bf y}}

\newcommand{\bfR}{{\bf R}}
\newcommand{\bfA}{{\bf A}}
\newcommand{\bfF}{{\bf F}}
\newcommand{\bfP}{{\bf P}}
\newcommand{\bfM}{{\bf M}}
\newcommand{\bfV}{{\bf V''}}

\newcommand{\bfX}{{\bf X}}
\newcommand{\bfVtilde}{{\bf \tilde{V}''}}
\newcommand{\bfzero}{{\bf 0}}
\newcommand{\bfIdent}{{\bf 1}}
\newcommand{\bfyp}{{\bf{y}_{\shortparallel}}}

\newcommand{\bfypj}{{\bf{y}_{\shortparallel,\mathit{j}}}}
\newcommand{\bfypjj}{{\bf{y}_{\shortparallel,\mathit{j+1}}}}
\newcommand{\bfyo}{{\bf Y_{\bot}}}
\newcommand{\bfyoo}{{\bf Y_{\bot,0}}}
\newcommand{\bomega}{{\mbox{\boldmath$\Omega$}}}
\newcommand{\bomegatilde}{{\mbox{\boldmath$\tilde{\Omega}$}}}
\newcommand{\bomegasmall}{{\mbox{\boldmath$\omega$}}}
\def\mem#1#2#3{  \left\langle #1 \left\vert  #2 \right\vert #3 
\right\rangle   }

\newcommand{\Sorth}{{S}_{\bot}}
\DeclareMathOperator\erf{erf}

\title{Instanton Rate Constant Calculations
Close to and Above the Crossover Temperature}

\author{Sean McConnell$^*$, 
Johannes K\"{a}stner\thanks{Institute for Theoretical Chemistry, University of 
Stuttgart, Pfaffenwaldring 55, 70569 Stuttgart, Germany}}

\begin{document}

\date{}
\maketitle

\begin{abstract}
Canonical instanton theory is known to overestimate the rate constant
close to a system-dependent crossover temperature and is inapplicable above
that temperature.
We compare the accuracy of the reaction rate constants
calculated using recent semi-classical rate
expressions to those from canonical instanton theory. We show that rate
constants calculated purely from solving the stability matrix for the action
in degrees of freedom orthogonal to the instanton path is not applicable at
arbitrarily low temperatures and employ two methods to overcome this.
Furthermore, as a by-product of the developed methods, we derive a simple
correction to canonical instanton theory that can alleviate this known
overestimation of rate constants close to the crossover temperature. The
combined methods accurately reproduce the rate constants of the canonical
theory along the whole temperature range without the spurious overestimation
near the crossover temperature. We calculate and compare rate constants on
three different reactions: H in the M\"uller--Brown potential,
methylhydroxycarbene $\to$ acetaldehyde and H$_2$ + OH $\to$ H + H$_2$O.
\end{abstract}

\begin{wileykeywords}
Atom tunneling, reaction rate, instanton theory, low-temperature reactivity,
computational chemistry, software update
\end{wileykeywords}

\clearpage

\begin{figure}[h]
\centering
\colorbox{background-color}{
\fbox{
\begin{minipage}{0.9\textwidth}
\includegraphics[width=8cm]{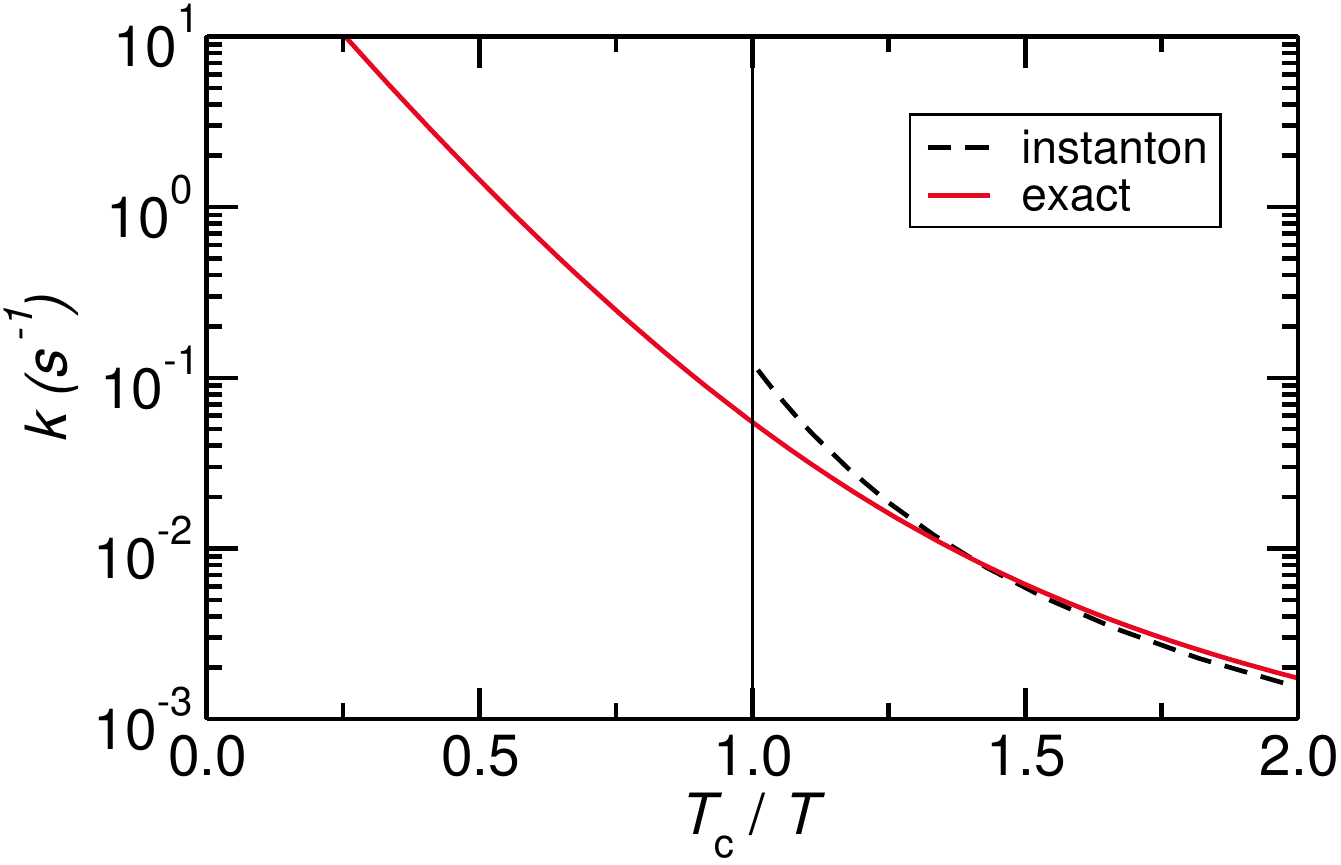}

Overestimation of the rate constant $k$ of an Eckart barrier by
instanton theory close to the crossover temperature $T_\text{c}$ compared
to the analytic solution. Our contribution proposes a technique to 
correct this overestimation which is computationally advantageous 
to the traditional canonical instanton approach.

\end{minipage}
}}
\end{figure}

  \makeatletter
  \renewcommand\@biblabel[1]{#1.}
  \makeatother

\bibliographystyle{myapsrev}

\renewcommand{\baselinestretch}{1.5}
\normalsize

\clearpage

\section{\sffamily \Large Introduction}

Semiclassical instanton theory is a well-established technique to calculate
reaction rate constants including quantum tunneling of atoms.\cite{kae14,mei16} In its most
common formulation, it is inapplicable above a system-dependent crossover
temperature $T_\text{c}=\frac{\hbar|\omega_b|}{2\pi k_\text{B}}$
 and is known to overestimate rate constants close to,
but below $T_\text{c}$. 
Here
$|\omega_b|$ is the absolute value of the imaginary frequency at the transition
state, and $k_\text{B}$ is the Boltzmann constant. Moreover, in terms of the 
computational demands, it requires the diagonalisation of a matrix of 
dimension $NP$, where $N$ is three times the number of atoms in the reaction 
and $P$ is the number of control points (images) of the instanton.

Recently, a new formulation of instanton theory has been postulated\cite{Kryvohuz2011,Kryvohuz2012,Kryvohuz2013}, which,
with our modifications, can overcome each of these problems while retaining good 
agreement with the canonical theory.

This implementation of instanton theory sits among others based on either evaluating 
the imaginary part of the free energy or representing the rate in terms of a 
flux through a surface.\cite{Langer1967,Langer1969,
Miller1975,Callan1977,Coleman1977,gil77,Affleck1981,col88,
Voth1989,han90,Benderskii1992,Cao1996,
Siebrand1999,Richardson2009,Goumans2010,Althorpe2011,Rommel2011,Kaestner2013} The
evaluation of free energies as a pathway to obtaining rate constants can be
traced back to the work in non-hermitian quantum mechanics in predicting the
decay of distinct quantum states by Gamow.\cite{Gamow1928} The utility of
these $Im(F)$/flux based methods in accounting for tunnelling lies in their applicability
to systems with many degrees of freedom. Though more rigorous techniques exist
for obtaining rate constants, e.g. wave-packet dynamics, \cite{Garraway1995}
these approaches are computationally prohibitive for all but the simplest of
systems, besides which, a number of publications exist justifying 
the agreement between instanton theory and more computationally intensive 
methods, a selection of which we cite here.\cite{Andersson2009,Tudela2014}
  
We set out to compare the temperature dependent rate constants determined by
this new theory, with our modifications, to that of canonical instanton theory.\cite{Rommel2011} 
Our comparison involves 3 different systems: the M\"uller--Brown potential,
the unimolecular reaction of methylhydroxycarbene to acetaldehyde and the
bimolecular, gas-phase reaction of a hydrogen molecule with a hydroxyl
radical forming water and a hydrogen atom.
In terms of the theoretical approach, the main difference between canonical
instanton theory and this new
formulation\cite{Kryvohuz2011,Kryvohuz2012,Kryvohuz2013} lies in the
calculation of the instanton partition function and the treatment of the
Euclidean action therein. Where in canonical instanton theory the temporal
integration in the action is discretised and the partition function
approximated using Laplace's method, the theory used here splits the action
into classical and orthogonal fluctuations. The contribution to the partition
function due to these orthogonal fluctuations can be calculated by determining
the eigenvalues of a monodromy matrix. The primary motivation behind the chosen 
test reactions is to gain an appreciation for the efficacy of the method on 
higher dimensional systems. In higher dimensions, each mode of the eigenvalue spectrum 
of the monodromy matrix is not so clearly separable from the remaining modes and 
moreover, a naive approach to obtaining the eigenvalue spectrum may imply an 
instable instanton orbit. From a theoretical standpoint, these are not necessarily 
questions of great importance, yet in practical terms they have the potential to limit 
the adoption of these techniques by a broader community.

The recent formulation\cite{Kryvohuz2011,Kryvohuz2012,Kryvohuz2013} is
applicable over the full temperature range, above and below the crossover
temperature $T_\text{c}$.
The well known
problem\cite{Arnaldsson2007,Goumans2011} of canonical instanton theory, that
reaction rate constants are overestimated at temperatures close to, but less
than $T_\text{c}$, is also solved.
Of course, the applicability of instanton theory can also be extended above
$T_\text{c}$ by switching to a microcanonic
expression.\cite{Miller1975,ric16,Richardson2016,McConnell2016} 

In comparison to canonical instanton theory, another advantage of this recent 
formulation\cite{Kryvohuz2011,Kryvohuz2012,Kryvohuz2013} is a reduced computational
burden. 

At no stage in the calculation presented
here is it necessary to diagonalise a matrix with dimension greater than
$N$. This ensures an acceleration in the calculation of $\mathcal{O}(P^2)$
over canonical instanton theory, for this reason, we will refer to this
approach henceforth as \emph{reduced instanton theory}.

Apart from the objective of method comparison, we seek to provide further
insights on the formulation and implementation of the theory.  To that end, we
will show empirically that contributions to the rate constant from tunnelling
orthogonal to the instanton path cannot be reliably accounted for by usage of
a stability matrix \cite{Kleinert2009} at arbitrarily low temperatures. We
illustrate a few techniques which can circumvent this problem.

The paper is structured as follows. In section \ref{sec:theory} we 
start from the established rate equation and illustrate the 
features contained in our implementation. Unique to our approach is 
the determination of the rate of change of the tunnelling energy 
$E_b$, and some technical detail in accounting for oscillations 
orthogonal to the instanton path. Thereafter, in section 
\ref{sec:Results}, we compare the performance of the reduced instanton
theory to canonical instanton theory in the aforementioned test cases 
at temperatures below $T_\text{c}$, this is the main objective of this 
contribution. For rate constants above $T_\text{c}$, a comparison to 
canonical instanton theory is not possible, in this region we 
therefore compare to the classical theory and to the Bell and Eckart 
approximations. We discuss the results, potential issues and future 
directions in sections \ref{sec:disc} and \ref{sec:conc}. All 
equations are in atomic units 
$(\hbar=m_e=4\pi\epsilon_0=1,\,\,c=1/\alpha)$. The theory as implemented here 
is available in the latest update to the quantum chemistry program DL-FIND.\cite{Kaestner2009a}

\section{\sffamily \Large Theory}\label{sec:theory}

  The starting point in our examination of the theory of rates in the 
  reduced instanton approach is the derivation by
  Kryvohuz\cite{Kryvohuz2011,Kryvohuz2013} for temperatures above and 
  below $T_\text{c}$. A detailed introduction to the underlying theory 
  can be found in these references and will not be repeated here.

  In the results section, we use the following formula\cite{Kryvohuz2013} 
  for reaction rate constants at temperatures above $T_\text{c}$ (or,
  equivalently, for $\beta<\beta_\text{c}=1/k_\text{B} T_\text{c}$)
  
  \begin{align}
    \label{Kryv_rate_oTc}
    k=&\frac{1}{Q_r}\frac{e^{-\beta V_0}}{2\beta_\text{c} \sin\left(\beta/\beta_\text{c}\right)}
    \Delta\sqrt{\frac{\pi}{2}}\left(1+\erf\left(-\frac{\Delta}{\sqrt{2}}\right)\right)e^{\Delta^2/2},\nonumber\\
    \text{where,}&\nonumber\\
    \Delta=&\frac{\sqrt{\beta\beta_\text{c}}}{1}\left(\frac{\beta_\text{c}^2}{\beta^2}-1\right)\sqrt{-\frac{dE_b(\beta_\text{c})}{d\beta}}.
  \end{align}
  
  The symbols of \eqref{Kryv_rate_oTc} are the reactant state 
  partition function $Q_\text{r}$, the reciprocal temperature 
  $\beta=1/k_\text{B} T$, the potential at the barrier top $V_0$, 
  and the rate of change of the tunnelling energy 
  $E_b\equiv E_b(\beta)$; $0<E_b(\beta)<V_0:$ $\forall$ 
  $\beta>\beta_\text{c}$. We assume that the reactant state energy is $0$.
  
  Our implementation of this established theory differs in the 
  computation of certain components, specifically, the rate of change 
  of the tunnelling energy $E_b$ as a function of the inverse 
  temperature $\beta$ and in the technical aspects of 
  calculating the contribution to the partition function due to 
  the action from tunnelling orthogonal to the instanton path. We 
  explore how to calculate the action from 
  orthogonal oscillations in section \ref{sec:Action_form} before examining 
  a simple technique to approximate $dE_b/d\beta$.

  In the reduced instanton theory, the rate constant at temperatures below
  $T_\text{c}$ is given by\citep{Kryvohuz2011}

  \begin{equation}\label{Kryv_rate}
    k=\frac{1}{2Q_\text{r}}\sqrt{\frac{-1}{2\pi}\left(\frac{dE_b}{d\beta}+\frac{d^2\sigma}{d\beta^2}\right)}
    e^{-S_\text{E}-\sigma}
    \times\left(1+\erf\left(\frac{\left(V_0-E_b\right)}
    {\sqrt{2}}\left(-\frac{dE_b}{d\beta}-\frac{d^2\sigma}{d\beta^2}\right)^{-1/2}\right) \right).
  \end{equation}

	We do not include zero-point corrections to $V_0$ or $E_b$ in \eqref{Kryv_rate} 
	as these corrections cancel near $T_c$ or become irrelevant at low 
	temperatures due to the term $dE_b/d\beta+d^2\sigma/d\beta^2$ rapidly 
	approaching zero.   
  New terms appearing in \eqref{Kryv_rate} that are absent from canonical 
  instanton theory are the classical action $S_\text{E}$ 
  and the contribution to the instanton partition 
  function by fluctuations orthogonal to the instanton trajectory 
  $\sigma$.
    
  It is useful to define the euclidean action $S_\text{E}$ as an integral 
  in imaginary time along an arbitrary path $\bfy$ in the 
  inverted, multidimensional potential $V(\bfy(\tau))$. In 
  mass-weighted coordinates it is given by
  \begin{equation}\label{ActionS}
    S_\text{E}\left[\bfy\right]=\int_{0}^\beta \left(\frac{1}{2}
    \left(\frac{d\bfy}{d\tau}\right)^2+V(\bfy(\tau))\right)d\tau.
  \end{equation}  
  The instanton path is a closed path that
  satisfies $\delta S_\text{E}\left[\bfy\right]=0$. Its length is 
  non-zero for all $\beta>\beta_\text{c}$. In our implementation, the vector 
  corresponding to the unstable mode at the transition state is used 
  as a starting guess for the instanton path at some $T\lesssim T_\text{c}$. 
  This trajectory is then optimised to meet the condition 
  $\delta S_\text{E}\left[\bfy\right]=0$, 
  we denote the coordinates of a point on this optimised path at 
  imaginary time $\tau$ by $\bfyp(\tau)$. At lower 
  temperatures, the optimised path at the previous higher temperature 
  is used as the new guess path and is re-optimised.
  \cite{Rommel2011a,Rommel2011}
      
  Unique to the reduced instanton approach is the formulation of $S_\text{E}$
  as well as explicitly requiring $dE_b/d\beta$. We first treat the 
  formulation of $S_\text{E}$ and then elaborate on an approximation for 
  $dE_b/d\beta$.     
  \subsection{\sffamily \large Action formulation\label{sec:Action_form}}

    Once the optimisation condition has been reached, the instanton action $S$ can
    be expanded into a classical component parallel to the instanton trajectory 
    and a component accounting for fluctuations orthogonal to the instanton
    trajectory.\cite{Gutzwiller,Kleinert2009}

    \begin{align}\label{Kryv_action}
    	S=&S_\text{E}+\Sorth,\nonumber\\
      S_\text{E}=&\int_{0}^{\beta}\left(\frac{1}{2}
      \left(\frac{d}{d\tau}\delta\bfyp\right)^2+
      V(\bfyp(\tau))\right)d\tau,\nonumber\\
      \Sorth=&\int_{0}^{\beta}\frac{1}{2}
      \left(\frac{d}{d\tau}\delta\bfyo\right)^2d\tau
      +\int_{0}^{\beta}\frac{1}{2}\delta\bfyo(\tau)^T
      \cdot\bfVtilde(\bfyp(\tau))\cdot\delta\bfyo(\tau)d\tau.
    \end{align}

    In \eqref{Kryv_action} and henceforth, a co-moving 
    coordinate system is used: 
    $\bfyp(\tau) \in \mathbb{R}^{N\times 1}$ and     
    $\bfyo(\tau)\in \mathbb{R}^{N\times(N-1)}$, 
    where $N$ is the dimension of the system. $\bfyo(\tau)$ is 
    a matrix of the other $N-1$ vectors orthogonal to $\bfyp(\tau+d\tau)-\bfyp(\tau)$. The fluctuations $\delta\bfyp(\tau)$ and $\delta\bfyo(\tau)$ are
    respectively, the scalar displacement along the path and an $N-1$ vector of displacements 
    orthogonal to the path. $\bfVtilde(\bfyp(\tau))\in \mathbb{R}^{(N-1)\times(N-1)}$ is 
    a reduced, rotated hessian, obtained by projecting $\bfV(\bfyp(\tau))$ onto the 
    $\bfyo(\tau)$ basis.
    
    The instanton path is discretised into $P$ images. The integrals in 
    \eqref{Kryv_action} are thus transformed into sums, 
    therefore $d\tau\to\Delta\tau=\beta/P$. This has the further 
    consequence that as temperature decreases, images accumulate 
    near the ends of the instanton path, 
    leaving the region around the transition state less well modelled.
      
    The integral $S_\text{E}$ of \eqref{Kryv_action} is 
    discretised into a Riemann sum:
    \begin{equation}\label{ActionS_disc}
      S_\text{E}=\beta\sum_{j=1}^P\left(\frac{P}{2\beta^2}|
      \bfypjj-\bfypj|^2+\frac{V(\bfypj)}{P}
      \right),
    \end{equation}      
    moreover, $e^{\sigma}$
    is the contribution to the partition function from fluctuations orthogonal
    to the instanton path 

    \begin{equation}\label{sigma_def}
      e^{-\sigma}=\prod_{i=1}^{N-1}\int d\,\delta\bfyoo^{(i)}
      \int_{\bfyo(0)=\bfyoo}^{\bfyo(\beta)=\bfyoo}
      \mathcal{D}[\delta\bfyo(\tau)]e^{-\Sorth},
    \end{equation}
    where the superscript $(i)$ indicates the $i-\text{th}$ vector 
    of the stationary basis $\bfyoo$.
    Depending upon which temperature regime we are in, either 
    $T\lesssim T_\text{c}$ or $T\ll T_\text{c}$, we make use of, respectively, the 
    stability matrix differential equation or frequency averaging to 
    determine $\sigma$. We also present in the results a third method 
    for calculating $\sigma$ known as eigenvalue tracing, an explanation
    of this method is given elsewhere\cite{McConnell2016} and we present
    it here for the purpose of comparison.
    \subsubsection{\sffamily \normalsize The stability matrix differential equation \label{sec:stab_mat_de}}

      A solution for $\sigma$ is found by solving the stability matrix 
      differential equation\cite{Kleinert2009,Schwidder2013}
      \begin{align}\label{GYeq}
        \frac{d}{d\tau}\bfR(\tau)+\bfF(\tau)\cdot\bfR(\tau)&=0;
        \,\,\bfR(0)=\bfIdent ,\nonumber\\
        \bfR(\tau)&=\left(\begin{array}{cc}
        \bfA_1(\tau) & \bfA_2(\tau)\\
        \dot{\bfA}_1(\tau) & \dot{\bfA}_2(\tau)
        \end{array}\right),
      \end{align}    
      where
      \begin{align}\label{GYF}
        \bfF(\tau)&=\left(\begin{array}{cc}
        \bfzero & -\bfIdent\\
        \bfV(\tau) & \bfzero
        \end{array}\right).
      \end{align}  
      The stability matrix differential \eqref{GYeq}, when solving for 
      $\bfA_1$ or $\bfA_2$, is known as the 
      the  Gel'fand-Yaglom equation.\cite{Gelfand1960}      
        
      The stability parameters (eigenvalues) $u_i$ of a matrix $\bfM$, where 
      $\bfR(\beta)=e^{\bfM}$, are sought. These uniquely 
      determine $\sigma$
      \begin{equation}\label{sigma}
        \sigma=\sum_{i=1}^{\mathfrak{N}}
        \ln\left(2 \sinh \frac{u_i}{2}\right);\,\,
        \mathfrak{N}=dim(\bfM)/2.
      \end{equation}    
      The dimension of $\bfF$, $\bfR$ and $\bfM$ is either $2N$ 
      or $2(N-1)$. In the former case, two extra zeros appear in the 
      eigenvalue spectrum of $\bfM$, these eigenvalues are excluded 
      from the sum in \eqref{sigma}. In the latter, we must 
      use a reduced, rotated potential Hessian $\bfVtilde(\tau)$. It 
      is constructed as follows.

      At each image we solve the eigenvalue equation
      $\bfV(\tau)\cdot\bfX(\tau)=\bomegasmall^2(\tau)\bfX(\tau)$ to 
      find the eigenvectors $\bfX(\tau)$ of the full potential 
      Hessian. The diagonal matrix $\bomegasmall^2(\tau)$ contains the force 
      constants. These eigenvectors are projected on to the basis 
      $\bfyo$, producing the projector matrix $\bfP(\tau)=
      \bfX(\tau)\cdot\bfyo(\tau)$. We then determine $\bfVtilde$ 
      according to:
      \begin{equation}\label{Hess_reduc}
        \bfVtilde_{N-1\times N-1}(\tau)=\bfP_{N-1\times N}^T(\tau)
        \cdot\bomegasmall(\tau)^2\cdot\bfP_{N \times N-1}(\tau).
      \end{equation}
 
      To begin, we choose
      an arbitrary starting image on the instanton path which is 
      assumed to correspond to $\tau=0$. At this image, an initial 
      guess basis is supplied to a Gram-Schmidt algorithm. For this,
      all but one of the eigenvectors $\bfX(0)$ and the normalised 
      vector $\bfyp(\Delta\tau)-\bfyp(0)$ are used 
      as the initial guess of the basis for $\bfyo(0)$. The 
      excluded eigenvector is the one that has the largest 
      projection onto $\bfyp(\Delta\tau)-\bfyp(0)$. Our implementation of the Gram-Schmidt
      algorithm ensures that the normalised vector $\bfyp(\tau+\Delta\tau)-\bfyp(\tau)$ is 
      always a part of the orthogonal co-moving basis at every image. 
      For subsequent co-moving bases $\bfyo(\tau)$, the new guess basis 
      again consists of the corresponding $\bfyp(\tau+\Delta\tau)-\bfyp(\tau)$, yet the guess 
      for the remaining $N-1$ basis vectors are the eigenvectors of the 
      Hessian at the previous image. In this way, our co-moving basis 
      retains a maximal degree of coherence which 
      makes possible the comparison of the eigenvectors of $\bfVtilde$ of 
      neighbouring images and permits the tracing of the
      eigenvalues of $\bfVtilde$ along the instanton path.
      
      The ansatz for eigenvalue tracing lies in the assumption that the 
      stability parameters $u_i$ of \eqref{sigma} can be interpreted as 
      frequencies orthogonal to, and averaged along, the instanton path. 
      The coherent nature of the proposed co-moving basis means that each 
      eigenvector of one particular Hessian is almost parallel to one (and only 
      one) eigenvector of a neighbouring Hessian. We can then ascribe the 
      eigenvalues of these corresponding eigenvectors to the same orthogonal 
      mode, the average frequency of this mode along the path gives the 
      required stability parameter.\cite{McConnell2016}
      
      Directly solving the stability matrix is a 
      reliable technique in a temperature range near, but below 
      the crossover temperature, where instanton paths are short. The 
      range of applicability of the stability matrix as a method 
      for obtaining $\sigma$ will be analysed in the discussion.
    \subsubsection{\sffamily \normalsize Frequency averaging}

      Below a certain system-dependent temperature, the 
      stability matrix method is unable to reliably calculate 
      $\sigma$. In the low temperature limit the eigenvalues 
      $u_i$, of \eqref{sigma}, exhibit an increasingly 
      linear dependence on $\beta$. We thus rewrite 
      $u_i=\beta\omega_{\bot,i}$, hence \eqref{sigma} 
      becomes
      \begin{equation}\label{sigma_lowT}
        \lim_{\beta\to\infty}\sigma\sim\sum_{i=1}^{\mathfrak{N}}
        \frac{\beta\bar{\omega}_{\bot,i}}{2},
      \end{equation}
      where $\mathfrak{N}$ is set to $N-1$.
      The set of frequencies $\bar{\omega}_{\bot}$ are \emph{average} 
      frequencies of the reduced, rotated Hessians $\bfVtilde$. 
      By analyzing \eqref{Kryv_action}, the following derivations 
      justify its usage in the low temperature regime.
      The kinetic part of \eqref{Kryv_action} can be partially integrated
      \begin{align}
        \int_{0}^{\beta}\frac{1}{2}
        \left(\frac{d}{d\tau}\delta\bfyo\right)^2d\tau=&\frac{1}{2}
        \left.\frac{\delta\bfyo^T}{d\tau}\cdot\delta\bfyo(\tau)\right|_{0}^{\beta}-
        \int_{0}^{\beta}\frac{1}{2}\delta\bfyo(\tau)^T\cdot
        {\frac{d^2}{d\tau^2}\delta\bfyo}d\tau.
      \end{align}
      The first term on the right is zero due to the boundary 
      conditions $\delta\bfyo(0)=\delta\bfyo(\beta)$, this is the condition of a 
      closed instanton path. \Eqref{Kryv_action} can be rewritten
      \begin{align}\label{Kryv_action_ave_freq}
        \Sorth=&\int_{0}^{\beta}\delta\bfyo(\tau)^T\cdot\left(-\frac{1}{2}
        \frac{d^2}{d\tau^2}+\frac{1}{2} 
        \bfVtilde(\bfyp(\tau))\right)\cdot\delta\bfyo(\tau)d\tau.
      \end{align}
      The term in brackets in \eqref{Kryv_action_ave_freq} 
      is the differential equation for uncoupled, quantum harmonic oscillators.
      This can be replaced with its diagonal eigenvalue matrix:
      \begin{align}\label{Kryv_action_QHO}
        \Sorth=&\int_{0}^{\beta}\delta\bfyo(\tau)^T\cdot
        \frac{\bomegatilde(\tau)}{2}\cdot\delta\bfyo(\tau)d\tau.
      \end{align}    
      In the limit $\beta\to\infty$, only the ground states of the 
      uncoupled quantum harmonic oscillators make a significant 
      contribution to the partition function, thus the elements of 
      $\bomega(\tau)$ are 
      $\bomega(\tau)_{nm}=\delta_{nm}\omega_{nm}(\tau)$.
      Discretising the integral in \eqref{Kryv_action_QHO} 
      one can rewrite \eqref{sigma_def}
      \begin{equation}\label{sigma_def_2}
        e^{\sigma}=\prod_{i=1}^{\mathfrak{N}}\int d\,\delta\bfyoo^{(i)}\Bigg[\delta\bfyoo^{(i)T}\cdot e^{\sum_{j=1}^{P}\left(
        \frac{\bomegatilde^{(j)}}{2}\right)\Delta\tau}
        \cdot\delta\bfyoo^{(i)}\Bigg]
      \end{equation}
      We are left with the definition of an operator trace. This 
      formulation for $\sigma$ has been used elsewhere.
      \cite{Kryvohuz2013}
      \begin{align}\label{sigma_derivation}
        e^{\sigma}=&\int d\,\delta\bfyoo \mem{\delta\bfyoo}
        {e^{\sum_{j=1}^{P}\Delta\tau
        \frac{\bomegatilde^{(j)}}{2}}}{\delta\bfyoo},\nonumber\\
        \sigma\sim&\frac{\beta}{2P}\sum_{j=1}^{P/2}
        \mathbf{tr}\left(\bomegatilde^{(j)}\right).
      \end{align}  
      We see that \eqref{sigma_derivation} corresponds 
      exactly to \eqref{sigma_lowT} given the trace is 
      conducted over $\mathfrak{N}$ dimensions.
      As with the stability matrix method, either full-dimensional 
      Hessians or the reduced, rotated Hessians can be used to find 
      $\sigma$. Using the full-dimensional Hessians means 
      $\mathfrak{N}=N$, in which case one adds a correction to  
      \eqref{sigma_derivation}:
      \begin{equation}\label{sigma_derivation_2}
        \sigma\sim\frac{\beta}{2P}\Re{\left[\sum_{j=1}^{P/2}
        \mathbf{tr}\left(\bomega^{(j)}\right)-
        \sqrt{\mem{\bfypj-\bfypjj}{\bfV_j}{\bfypj-\bfypjj}}\right]},    
      \end{equation}
      This removes any contribution to $\sigma$ from oscillations 
      parallel to the instanton path. \Eqref{sigma_derivation} (or 
      \eqref{sigma_derivation_2}) is a convenient form for 
      $\sigma$, it is simply a sum of all the eigenvalues from the 
      Hessians at each image. It should also be noted that for 
      computational purposes, one should take only the real part of the 
      RHS of \eqref{sigma_derivation_2} because, for a small $P$, 
      spurious, complex contributions to $\sigma$ tend to increase.

      As opposed to unimolecular reactions, 
      calculation of both reactant and instanton partition functions in 
      bimolecular reactions demand a different approach to the treatment
      of very small frequencies at low temperatures. In unimolecular 
      reactions, since the images of the instanton path accumulate near 
      the reactant state, the conditions imposed on the treatment of 
      small frequencies in the reactant state partition function can be 
      equally applied to small frequencies in the instanton partition 
      function. In bimolecular reactions, at low temperatures, the 
      images of the instanton path may accumulate near a pre-reactive 
      minimum, the coordinate and potential landscape of which is likely
      very different when compared with the reactant state of the 
      separated system. The conditions for eliminating small 
      frequencies from the instanton partition thus need modifying. In 
      the calculations shown for the reaction H$_2$ + OH $\to$ H + 
      H$_2$O in the frequency averaging scheme, we omit from 
      \eqref{sigma_derivation} and \eqref{sigma_derivation_2} those 
      frequencies where $\beta\bar{\omega}_{\bot,i}/2<\sinh^{-1}(1/2)$. 
      This is justified since 
      \begin{align}\label{low_freq_cond}
        \left.
        \begin{array}{ccc}
          \mathrm{sgn}(\log[2\sinh(x)]) &=& \mathrm{sgn}(x)\\
          \log[2\sinh(x)]&\sim& x
        \end{array}\right\rbrace&\forall:\, x>\sinh^{-1}(1/2).
      \end{align}  
      The second condition of \eqref{low_freq_cond} is in any 
      case the main assumption behind the frequency averaging approach.
      It is natural to ask at which $\beta$ the stability matrix 
      approach should give way to frequency averaging. The reliability 
      of the stability matrix approach depends strongly on the length 
      of the path, which, depending on the potential, can change drastically 
      between two similar temperatures. From \eqref{low_freq_cond} 
      we know that frequency averaging 
      is maximally inaccurate when $\beta=\beta_c$, we can ask the more instructive 
      question ``what is the upper bound on the relative uncertainty $\delta$ in 
      $\sigma$ when calculated using frequency averaging?'' This upper bound is 
      given by 

			\begin{equation}\label{eq:switching_beta}
				\delta(\beta>\beta_c)<\frac{\sum_{i=1}^\mathfrak{N}\log\left(1-e^{-\beta_c\bar{\omega}_{\bot,i}^{\mathrm{TS}}}\right)}{\sum_{i=1}^\mathfrak{N}\beta_c\bar{\omega}_{\bot,i}^{\mathrm{TS}}/2}.
			\end{equation}

			As an example, for the reaction methylhydroxycarbene $\to$ acetaldehyde 
			treated in the results we get $\delta(\beta_c)=4.97\%$. 
      
      Furthermore, in the version implemented in DL-FIND, the zero eigenvalues of $\bfM(\beta)$ 
		  (identified by projection on to the path) are displayed in the output. 
		  The user may, independently of \eqref{eq:switching_beta}, decide at which 
		  temperature these zero eigenvalues are no longer negligible and 
		  hence determine the temperature at which the calculation of $\sigma$ 
		  using the eigenvalues of the stability matrix are no longer reliable. 
      
      Having defined how the action enters into \eqref{Kryv_rate} 
      we now explore the methods available for deriving $dE_b/d\beta$.

  \subsection{\sffamily \large Temperature dependence of the tunnelling energy}
  \label{sec:dedb}
    
    The most accurate approach to determine $dE_b/d\beta$ is 
    a finite difference procedure. To this end, the instanton path is 
    first  optimised at two nearby temperatures. The endpoints of the 
    path lie at a particular energy on the potential surface, this 
    energy is the tunnelling energy $E_b$. A simple ratio between 
    $\Delta E_b$ and $\Delta\beta$ for the two paths is used
    to approximate $dE_b/d\beta$.
    
    Re-optimising the instanton path for small temperature changes 
    around the desired temperature can be time consuming. Instead we 
    provide an approximate formula for estimating $dE_b/d\beta$ which 
    is sufficiently accurate for all temperatures $T<T_\text{c}$.
    
    From the stationary condition for locating instantons, 
    $\delta S_\text{E}=0$, one can derive the following relation
    \cite{Rommel2011,Kleinert2009}
    \begin{equation}\label{dEdB1}
      \beta =\int_{s_a(E_b)}^{s_b(E_b)}\sqrt{\frac{2}{V(\bfy(s))-E_b}}
      \left|\frac{d\bfy(s)}{ds}\right|ds,
    \end{equation}
    where $V(\bfy(s_a))=V(\bfy(s_b))=E_b$. For all $s \in 
    \mathbb{R}|s_a<s<s_b$, the tunnelling energy is 
    less than the potential energy. We approximate 
    the integral at this point by expanding the potential in a Taylor series. 
    The expansion is fixed around the reactant and product state coordinates, ensuring 
    the proper behaviour of $dE_b/d\beta$ at both limits $\beta\to\infty$ or 
    $\beta\to\beta_\text{c}$. Moreover, \Eqref{dEdB1} remains 
    unchanged if the radicand is an absolute value
    \begin{eqnarray}\label{dEdB1.1}
      \beta \sim& \bigint_{s_a(E_b)}^{s_b(E_b)}
      \sqrt{\frac{2}{\left|(V_\text{RS} - E_b) + \bfy(s)^T\cdot\frac{\bfV_\text{RS}}{2}\cdot
      \bfy(s)\right|}}
      \left|\frac{d\bfy(s)}{ds}\right|ds.\nonumber\\
      +&\bigint_{s_a(E_b)}^{s_b(E_b)}
      \sqrt{\frac{2}{\left|(V_\text{PS} - E_b) + \bfy(s)^T\cdot\frac{\bfV_\text{PS}}{2}\cdot
      \bfy(s)\right|}}
      \left|\frac{d\bfy(s)}{ds}\right|ds.
    \end{eqnarray}
    If we are dealing with asymmetric potentials, the second part of \eqref{dEdB1.1} 
    will always be much smaller than the first part.
    Because the limits of the integral themselves depend on the 
    tunnelling energy one would ordinarily 
    use the Leibniz integral rule to find the derivative with respect to $E_b$. 
    However, we retain only the following term in the Leibniz integral rule
    \begin{align}\label{dEdB3}
      \frac{d\beta}{dE_b}\sim&-\sqrt{\frac{2}
      {\left|\left(V_\text{RS}-E_b\right)+
      \bfy(s_a)^T\cdot\frac{\bfV_\text{RS}}{2}\cdot
      \bfy(s_a)\right|}}\times\frac{d\left|\bfy\left(s_a\right)\right|}{dE_b},
    \end{align}
		we may thereby provide a good approximation to full Leibniz rule derivation of $d\beta/dE_b$ 
		for most tunnelling energies and remain closer the finite difference estimate for 
		tunnelling energies approaching $V_\text{RS}$.   
		The vector $\bfy(s_a)$ connects the 
    expansion coordinate to that image on the instanton path nearest 
    the reactant state.  Furthermore, we may freely choose the parameterisation, and set 
    $s_a=\left|\bfy\left(s_a\right)\right|$ and $\left|\frac{d\bfy(s_a)}{ds}\right|=1$ for all 						tunneling energies.
    We make use again of the Taylor series approximation to find a 
    closed form for $d\left|\bfy\left(s_a\right)\right|/dE_b$.
    \begin{align}\label{dEdB4}
      E_b-V_\text{RS}&\sim \bfy(s_a)^T\cdot\frac{\bfV_\text{RS}}{2}
      \cdot\bfy(s_a),\nonumber\\
      E_b-V_\text{RS}&\sim \left|\bfy(s_a)\right|^2
      \left(\hat{\bfy}(s_a)^T\cdot\frac{\bfV_\text{RS}}{2}
      \cdot\hat{\bfy}(s_a)\right),\nonumber\\
      \frac{dE_b}{d\left|\bfy(s_a)\right|}&\sim \left|\bfy(s_a)\right|
      \left(\hat{\bfy}(s_a)^T\cdot\bfV_\text{RS}\cdot\hat{\bfy}(s_a)\right).
    \end{align}
    Combining \eqref{dEdB3} and \eqref{dEdB4} provides reasonably 
    accurate and stable results for all temperatures below $T_\text{c}$.
    We note that this formulation is reasonable for asymmetric barriers because 
    the leading order contribution to the integral in Equations (\ref{dEdB1}) 
    and (\ref{dEdB1.1}) comes from a small 
    region near the reactant state, or equivalently, when $s\sim s_a$. For 
    nearly symmetric barriers, one should multiply Equations (\ref{dEdB1.1}) 
    and (\ref{dEdB3}) by a factor of 2. 
    
\begin{figure}
      \centering
			\includegraphics[width=8cm]{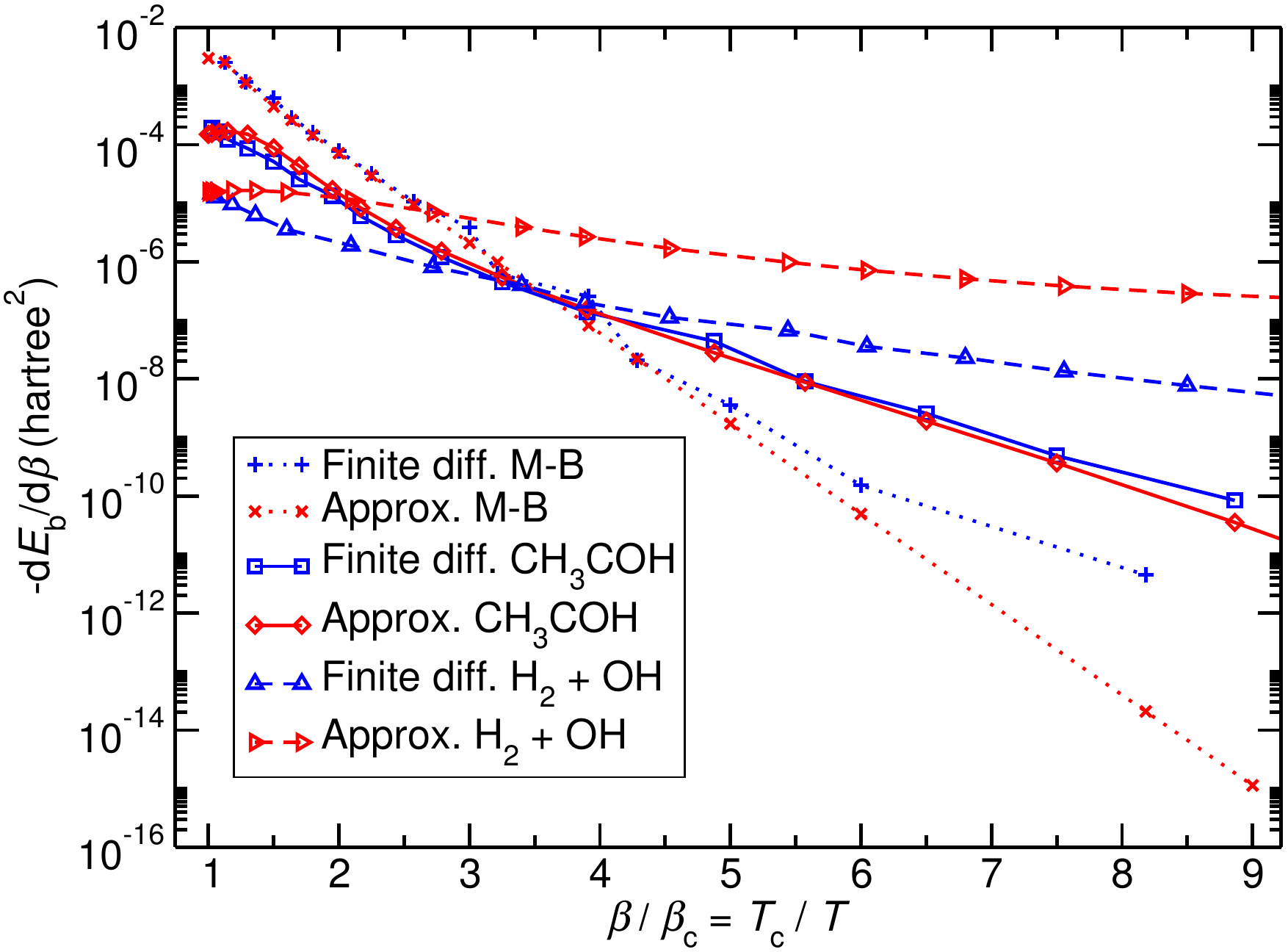}
      \caption{Temperature dependence of the tunnelling energy in the reaction of
        each of the reactions examined in section \ref{sec:Results}, red
         lines were calculated using \eqref{dEdB3}.  \label{fig:dedb}}
    \end{figure}

    As the temperature approaches the crossover temperature, a 
    more accurate method\cite{Kryvohuz2013} is employed to determine 
    $dE_b/d\beta$. Since all images of the instanton collapse to a 
    point when $T=T_\text{c}$, \eqref{ActionS} simplifies greatly, 
    i.e. the kinetic part is zero. Using the fact that 
    $d^2S_\text{E}/d\beta^2=dE_b/d\beta$ 
    and expanding the potential 
    in higher orders,\cite{Cao1996,Kryvohuz2013} one can modify 
    \eqref{dEdB3} for very short tunnelling paths in order to 
    continue calculating rate constants near $T_\text{c}$. 
    To accomplish this and to guarantee smoothness of $dE_b/d\beta$ 
    over all temperatures below $T_\text{c}$, we use a weighted combination 
    of $dE_b/d\beta$ calculated by expanding the potential around the 
    reactant state i.e. \Eqref{dEdB3} and the method expounded in
    \citet{Cao1996} and \citet{Kryvohuz2013} wherein paths are 
    represented as a fourier series, permitting a closed expression 
    for $dE_b/d\beta$ depending on anharmonic terms in the taylor 
    series for the potential around the transition state. With respect
    to the distance between the reactant state coordinate 
    and the coordinate of that image on the instanton path nearest the
    reactant state, the weighting of the method in \citet{Cao1996} and 
    \citet{Kryvohuz2013} increases with the inverse cube of this 
    distance whereas the weighting of \eqref{dEdB3} drops linearly with 
    this distance. A comparison is shown in \figref{fig:dedb} 
    between the finite difference and approximation approaches for the
    reactions examined in the results. 

    
    Clearly the approximation method for $dE_b/d\beta$ in the reaction 
    H$_2$ + OH $\to$ H + H$_2$O does not agree as nicely with the finite-difference 
    method as do the other two cases. 
    The potential landscape around the reactant state for the reaction on the M\"uller-Brown surface 
    and for methylhydroxycarbene, are quite deep and well approximated by a 
    harmonic potential (2$^\mathrm{nd}$ order Taylor series) 
    in comparison to the reaction H$_2$ + OH $\to$ H + H$_2$O. 
    The approximations used in \eqref{dEdB1.1} and \eqref{dEdB4} are made only 
    to second order. This suggests that $dE_b/d\beta$, when calculated by the given 
    approximation, and applied to reactions with shallow reactant states 
    or pre-reactive minima will not be well modelled except at temperatures close to $T_c$. 
    Fortunately, the rate constant is not overly sensitive to inaccuracies in $dE_b/d\beta$ since 
    it enters the rate equation under the square root sign.

  \subsection{\sffamily \large A correction to canonical instanton theory
    close to $T_\mathrm{c}$\label{sec:inst_corr}}
    As mentioned in the introduction, an endemic problem of canonical instanton
    theory is the overestimation of rate constants at temperatures approaching
    $T_\text{c}$. The derivation of rate \eqref{Kryv_rate} corrects this behaviour
    by ensuring those tunnelling paths with energies $E_b>V_0$ do not contribute
    to the rate calculation.
    
    A better understanding of this peculiarity  
    can be gained by examining the equation for the canonical instanton theory 
    rate $(k_\text{inst})$ and the expression for the flux $(f)$ over a barrier 
    used to determine the rate in \eqref{Kryv_rate}
    \begin{align}
    	\label{eq:flux}
    	k_\text{inst}&=\frac{2}{\beta}\mathfrak{I}\left(\log\left(Q\right)\right)\sim\frac{2}{\beta}\frac{\mathfrak{I}\left(Q_\text{inst}\right)}{Q_r},\nonumber\\
    	k&=\frac{f}{Q_r}=\frac{1}{2\pi Q_r}\int_{-\infty}^{V_0}e^{-\beta
          E} \sum_{k=1}^\infty\left(-1\right)^{k-1}e^{-k S_0(E)}\prod_{i=1}^{N}\frac{1}{2\sinh{ku_i/2}}dE,
    \end{align}
    
		where $Q_\text{inst}$ is the partition function of the instanton, $S_0$ 
		is the shortened action and the term
		under the sum is the cumulative reaction probability.\cite{Miller1975} At low 
	  temperatures it is sufficient to truncate the sum over $k$ to only the first term, 
	  furthermore, the term under the product should look familiar from \eqref{sigma}, thus 
	  we may simplify. 
	  \begin{equation}
    	\label{eq:flux2}
    	k=\frac{f}{Q_r}=\frac{1}{2\pi Q_r}e^{-\sigma}\int_{-\infty}^{V_0}e^{-\beta
          E}e^{-S_0(E)} dE,	  	
	  \end{equation}
	  
	  As explained in the appendix of Ref. \citenum{Kryvohuz2011}, 
		$S_0$ in \eqref{eq:flux} can be expanded to second order around $E=E_b$, which 
		results in an expression containing the error function, indeed this is the 
		term in the large braces in \eqref{Kryv_rate}.
		
		In canonical instanton theory, the
    partition function of the instanton $(Q_\text{inst})$ is expressed as an integral in configuration
    space over all possible closed paths. At temperatures approaching $T_\text{c}$ 
    an increasing number of closed paths possess an energy $E_b>V_0$. There is no way to 
    selectively remove these paths from the integral in canonical 
    instanton theory. This problem persists for all
    $T<T_\text{c}$, however, the contribution to the partition function from paths
    with $E_b>V_0$ as $T\to 0$ K becomes vanishingly small. When using the expression 
    for the flux, we can explicitly control which states contribute to 
    the rate expression by setting the 
    upper limit of the integral to $V_0$, this effectively truncates the Boltzmann 
    distribution for the state occupancy to zero for those states with energy 
    greater than the barrier height. This is a reasonable step to take, since those 
    states face no restriction to recrossing.
   
    If however we replace the upper limit of the integral in 
		\eqref{eq:flux} with $\infty$ the error function becomes 1, with the 
		rate expression simplifying to 
		\begin{equation}
			\label{Kryv_rate_noerf}
      k=\frac{1}{Q_\text{r}}\sqrt{\frac{-1}{2\pi}\left(\frac{dE_b}{d\beta}+\frac{d^2\sigma}{d\beta^2}\right)}
      e^{-S_\text{E}-\sigma}.		
		\end{equation}
		In this case, as was shown in a paper by \citet{Althorpe2011}, the canonical 
		instanton rate $k_\text{inst}$, and the rate $k$ determined by 
		\eqref{Kryv_rate_noerf} are equivalent and both would exhibit the same 
		overestimation of the rate near $T_c$.
        
    The error function in \eqref{Kryv_rate} takes on the role of a correction
    factor and may be used to modify the results of a rate calculation using
    canonical instanton theory ($k_\text{inst}$) in exactly the same way it 
    modifies the rate calculated using the flux over the barrier approach
    \begin{equation}\label{eq:inst_corr}
      k_\text{corr}=k_\text{inst}\frac{1}{2}
      \left(1+\erf\left(\frac{\left(V_0-E_b\right)}
           {\sqrt{2}}\left(-\frac{dE_b}{d\beta}\right)^{-1/2}\right)\right).
    \end{equation}
    At $T=T_\text{c}$ we have $E_b=V_0$ so that
    $k_\text{corr}=k_\text{inst}/2$. At low temperature, $E_b\ll V_0$ and the
    correction factor is almost unity, $k_\text{corr}\approx k_\text{inst}$. 

  \subsection{\sffamily \large Rate constant calculations with Bell and Eckart
    approximations}

We compare our instanton rate constants to one-dimensional tunneling
approximations and to rate constants obtained without tunneling. Even though
these approaches have been used in our group's previous
work,\cite{Goumans2011,alv14,lam16,son16,alv16,Meisner2016,lam17,kob17} they
have not yet been described in detail. In harmonic transition state theory, the
potential energy surface around the reactant and the transition
structure is approximated harmonically. The vibrational frequencies of
the reactant are denoted $\omega_{\text{RS},i}$ and those of the
transition state $\omega_{\text{TS},i}$. For a system with $D$
vibrational degrees of freedom, the rate constant in harmonic
transition state theory is
\begin{equation}
k_\text{HTST}(\beta)=\frac{1}{2\pi\beta\hbar}
\frac{Q_\text{rot,TS}}{Q_\text{rot,RS}}
\frac{Q_\text{trans,TS}}{Q_\text{trans,RS}}
\frac{\prod_{i=1}^D 2
  \sinh(\beta\hbar\omega_{\text{RS},i}/2)}
{\prod_{i=1}^{D-1} 2 \sinh(\beta\hbar\omega_{\text{TS},i}/2)}
\exp(-\beta V_0).
\label{eq:hqtst:kqvib}
\end{equation}
The rotational partition function is denoted
$Q_\text{rot}$ and the translational partition function
$Q_\text{trans}$. \Eqref{eq:hqtst:kqvib} can be used for uni-molecular (in
which case the ratio $Q_\text{trans,TS}/Q_\text{trans,RS}$ is unity) as well
as bimolecular rate constants. All vibrational degrees of freedom are treated
as quantum harmonic oscillators. Thus, \eqref{eq:hqtst:kqvib} includes effects
of the vibrational zero point energy in the harmonic approximation, but
neglects tunneling. Is is denoted as ``Classical + ZPE'' in the following.

In a coarse approximation, tunneling can be incorporated by using
a one-dimensional tunneling correction, i.e., by assuming that
tunneling happens only along the reaction coordinate and the
vibrational levels are not affected. Then $k_\text{Eckart}$, the rate
constant including tunneling with the Eckart approximation, as used in
the present work, is
\begin{equation}
  k_\text{Eckart}=k_\text{HTST} \kappa_\text{Eckart}
  \label{eq:2}
\end{equation}
where $\kappa_\text{Eckart}$ is the tunneling correction factor. It is
calculated as fraction of the quantum flux and the classical flux
through the barrier:
\begin{equation}
  \kappa=\frac{f_\text{quantum}}{f_\text{classical}}=\beta \exp(\beta V_\text{VA}) \int_0^\infty P(E) \exp(-\beta E) dE.
\end{equation}
Here, $V_\text{VA}$ is the vibrationally adiabatic barrier, i.e.  $V_0$
plus the difference in zero point energy between reactant and transition
state, $E$ given relative to the vibrationally adiabatic energy of the
reactant, and $P(E)$ the quantum mechanical transmission coefficient. For a
symmetric Eckart-shaped barrier with the same barrier frequency as the full
barrier and the same height as the vibrationally adiabatic barrier,
$P_\text{Eckart}(E)$ is given as\cite{eck30}
\begin{equation}
  P_\text{Eckart}(E)=\frac{\cosh(2a)-1}{\cosh(2a)-\cosh(d)}
  \label{eq:31}
\end{equation}
with
\begin{eqnarray}
a&=& \frac{2\pi\sqrt{V_\text{VA}E}}{\hbar\omega_b}\\
d&=& \frac{2\pi\sqrt{4V_\text{VA}^2-\hbar\omega_b/4}}{\hbar\omega_b}
\end{eqnarray}
This approach is denoted as Eckart approximation in the following. One
  could also use the asymmetric Eckart barrier instead of \eqref{eq:31}, but
  that would require an additional parameter and generally only marginally
  improves the accuracy.

Alternatively to the symmetric Eckart barrier, one can also approximate
tunneling by the transmission probability of a truncated parabolic barrier
with the barrier frequency $\omega_b$. Bell \cite{bel59} derived the tunneling
correction factor $\kappa_\text{Bell}(T)$ as the exact solution of the
permeability of a truncated parabolic barrier (2$^\text{nd}$ order polynomial):
\begin{equation}
  \kappa_\text{Bell}(T)=\frac{\pi u}{\sin(\pi u)} -
  u\exp(a-b)
  \left[\frac{1}{1-u}-\frac{\exp(-b)}{2-u}+\frac{\exp(-2b)}{3-u}-\ldots \right]
  \label{eq:hqtst:kappa:bell2}
\end{equation}
with $u=T_\text{c}/T$, $a=\beta V_\text{VA}$, and $b=V_\text{VA}/(k_\text{B}T_\text{c})$. Note
that $\kappa_\text{Bell}(T)$ is finite even for $T\rightarrow T_\text{c}$,
$T\rightarrow T_\text{c}/2$, \ldots . This approach
is denoted ``Bell'' in the following.

\section{\sffamily \Large Results}\label{sec:Results}

  We now apply the reduced instanton theory and the corrected
  canonical instanton theory to three different 
  systems:
  The two-dimensional M\"uller--Brown potential, the unimolecular reaction
  methylhydroxycarbene $\to$ acetaldehyde ($N=21$) and the bimolecular
  gas-phase reaction H$_2$ + OH $\to$ H + H$_2$O ($N=12$).  The objective is
  to make a comparison between the reduced instanton theory and canonical
  instanton theory but also to illustrate how
  the overestimation of rate constants near $T_\text{c}$ is  
  reduced by both this method and by using \eqref{eq:inst_corr}. In the
  temperature range over $T_\text{c}$ we compare with rate constants
  calculated using classical theories as well as the fitted one-dimensional
  potentials (Eckart and Bell).
    
  In this work, we compare methods rather than aiming at new insight into
  specific molecular systems. Therefore, we have chosen systems for which
  the potential energy and its derivatives can be calculated efficiently and
  without numerical noise. Instanton theory, however, is applicable
  efficiently to
  real-world applications with energies calculated on the fly as we and
  others have demonstrated frequently in the past.\cite{cha75,mil94,mil95,mil97,Siebrand1999, sme03,qia07,Andersson2009,
  gou10a,gou11,gou11b, Rommel2011a,Goumans2010,jon10,mei11,Goumans2011,ein11,rom12,
  Kryvohuz2012,Kaestner2013,alv14,kry14,lam16,son16,alv16,lam17,kob17}

  \begin{figure}[!htb]
      \includegraphics[height=0.3\textheight]
      {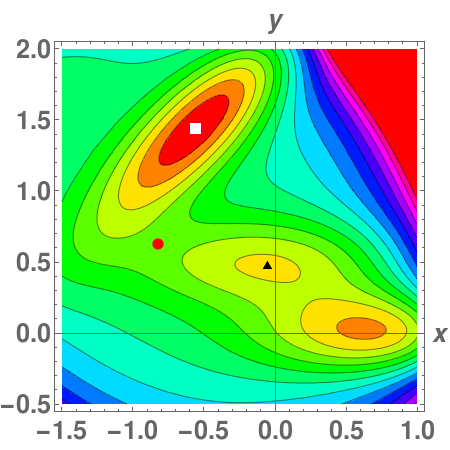}
      \caption{\label{fig:MBSurface}The M\"uller--Brown PES. The black triangle 
      represents the location of the reactant 
      state $(-0.050, 0.467)$, the white square is the global 
      minimum $(-0.558, 1.44)$ and the red disc is the transition state 
      $(-0.822, 0.624)$. The potential difference between the reactant and 
      transition state is 0.19 Hartree.}
  \end{figure}
  \begin{figure}[!htb]
		\includegraphics[width=8cm]{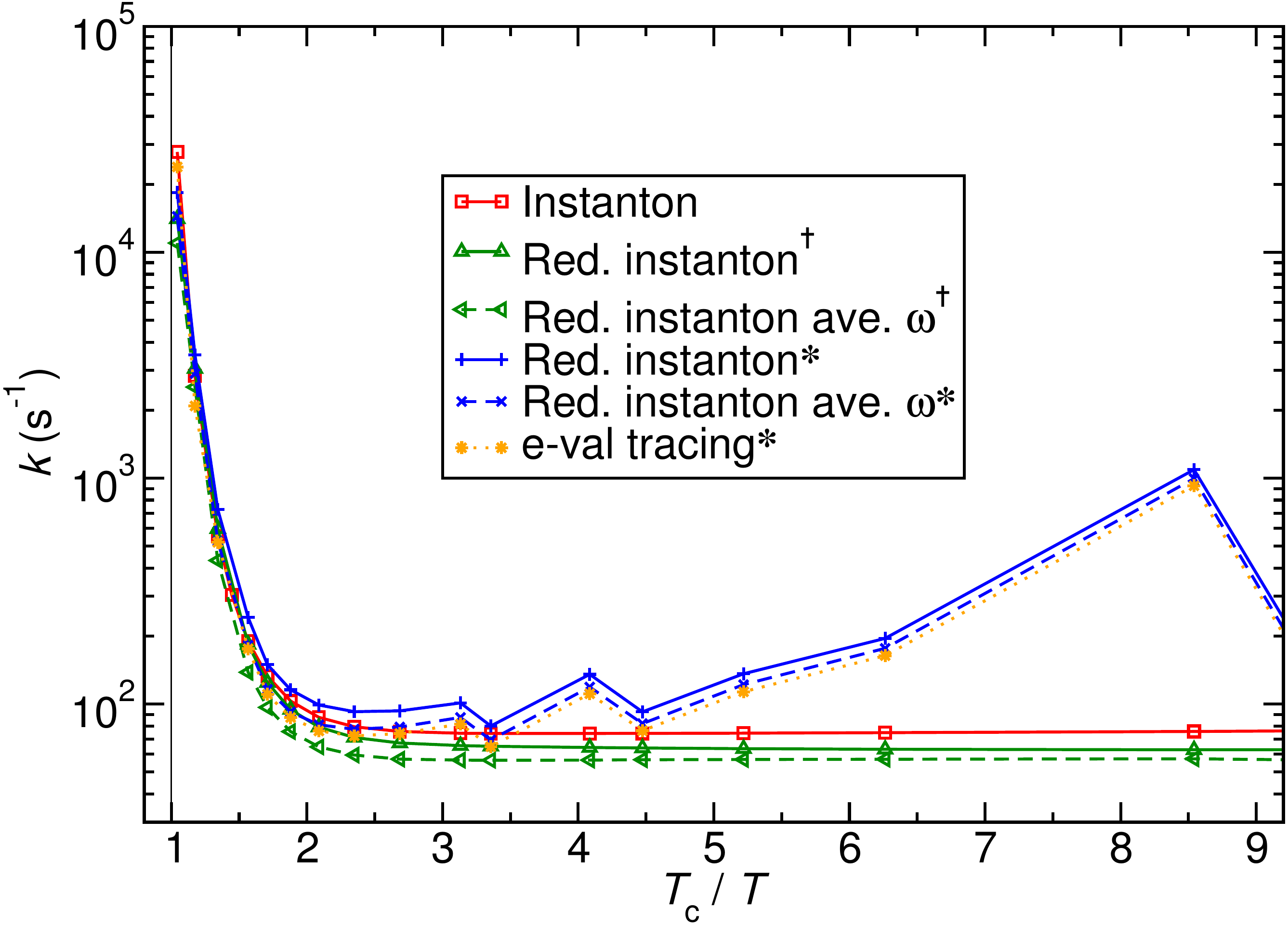}
		\includegraphics[width=8cm]{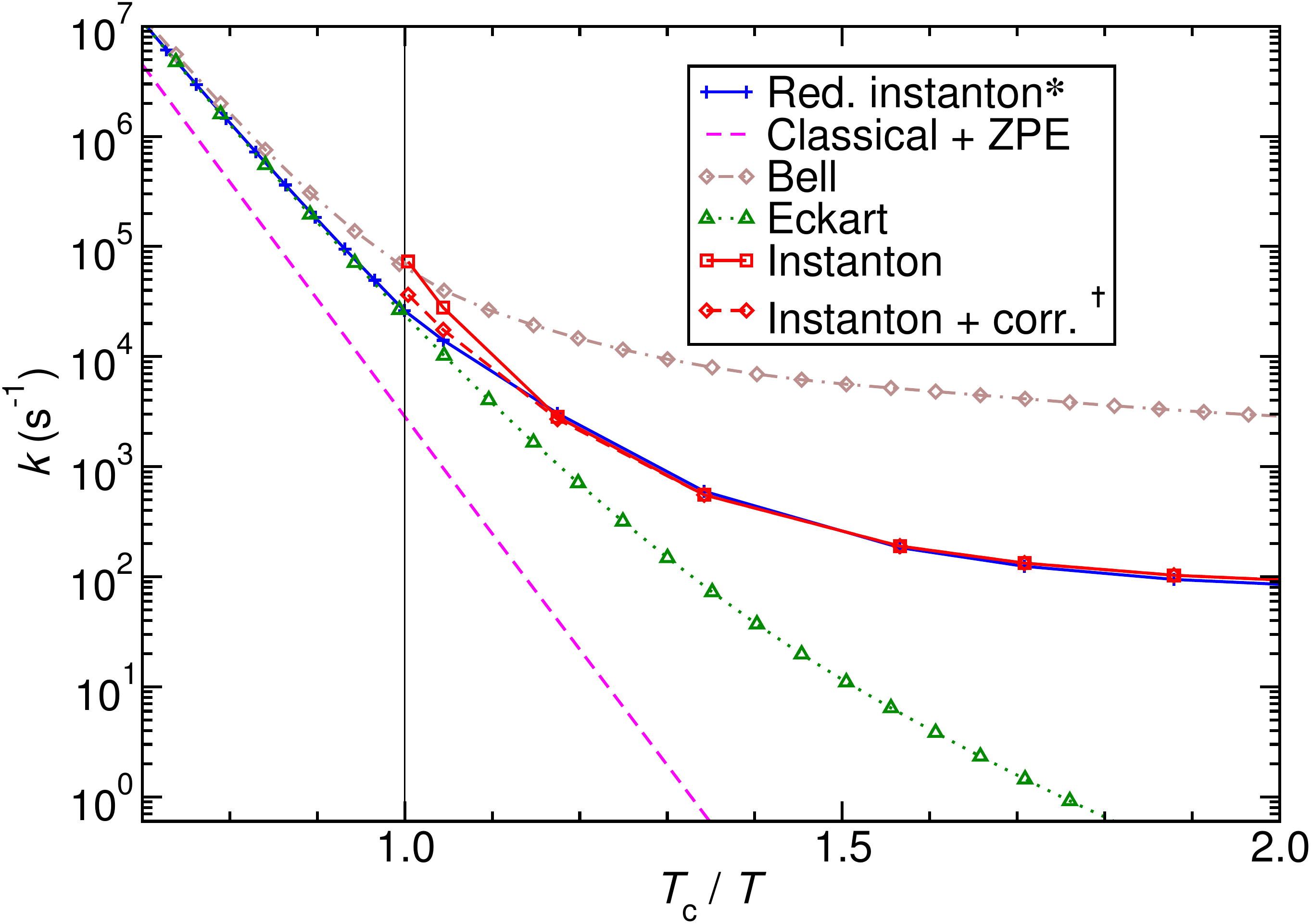}
    \caption{Rate constants for the M\"uller--Brown potential. Comparison
    between canonical instanton theory (red line) and reduced instanton theory
    (blue \& green lines). The symbol $\ast$ signifies $dE_b/d\beta$ was
    calculated using the finite difference method, $\dagger$ indicates
    $dE_b/d\beta$ was calculated using \eqref{dEdB3}. Red diamonds 
    refer to \eqref{eq:inst_corr}. The top graph
    shows temperatures below $T_\text{c}$, the bottom graph around and above
    $T_\text{c}$. Comparisons to classical transition state theory and the
    Bell and Eckart approximations are given.\label{fig:MB_Tc}}
  \end{figure} 
  \setcounter{subfigure}{0}

  \subsection{\sffamily \large The M\"uller--Brown surface}

    The M\"uller--Brown surface\cite{Mueller1979} 
    is a common ``toy model'' for testing reaction rate theories. We used the mass
    of a hydrogen atom and the minimum at the coordinate $(-0.050, 0.467)$ as the
    reactant state. This results in a potential energy barrier of 0.19 Hartree and
    a crossover temperature of 2207 K. The main advantage of testing on a two
    dimensional surface comes from the fact that, at each image, there exists only
    one, two-component vector orthogonal to the instanton path. The utility of
    this means that there should be no difference between the frequency-averaging
    approach, or the stability matrix method in the evaluation of $\sigma$ at low
    temperatures, except for small numerical errors arising from the second part
    of \eqref{sigma_derivation_2} which are caused by the discrete nature of the
    path.This is indeed the case, as can be seen in \figref{fig:MB_Tc}.

    Interestingly, due to the fact that the tunnelling energy and the 
    energy of the reactant state are numerically very close, the finite 
    difference method (blue and orange curves, \figref{fig:MB_Tc}) 
    to determine $dE_b/d\beta$ becomes unreliable. This is evidence of 
    the utility of the formula derived in section \ref{sec:dedb}.
    
    For those temperatures near and above $T_\text{c}$ 
    (\figref{fig:MB_Tc}) we can see that the reduced
    instanton theory performs favourably, 
    producing rate constants which smoothly transition towards those 
    found by the formula for rate constants above $T_\text{c}$. Also evident is
    the deviation in the rate constants calculated by canonical 
    instanton theory in this limit, it is clear there is an 
    overestimation of the rate constant near $T_\text{c}$ which can be
    partly corrected by \eqref{eq:inst_corr}.
  
  \subsection{\sffamily \large Unimolecular reaction: methylhydroxycarbene $\to$ acetaldehyde}
    The reaction of methylhydroxycarbene to acetaldehyde,
    CH$_3$COH $\rightarrow$ CH$_3$CHO, via tunnelling
    mechanisms has gained some attention recently.
    \cite{sch11a,Ley2012}$^,$\cite{Kaestner2013} 
    At low temperatures, although a lower barrier towards the
    formation of vinyl alcohol is present, the 
    formation of acetaldehyde is favoured due to the shorter tunnelling 
    path. This, and the fact that it is a reaction in many degrees of 
    freedom, make it a solid candidate to compare the canonical 
    and reduced instanton theories.
  
    For this work, a local potential energy surface was created by 
    training a neural network to fit the DFT potential used 
    previously\cite{Kaestner2013} in order to provide a fast and noise-free potential. 
    Possible inaccuracies in the fit 
    should not affect the comparisons performed here. The potential 
    energy barrier height resulted in 133~kJ~mol$^{-1}$
    and the crossover temperature was $T_\text{c}=461$ K.
  
\begin{figure}
      \includegraphics[width=8cm]{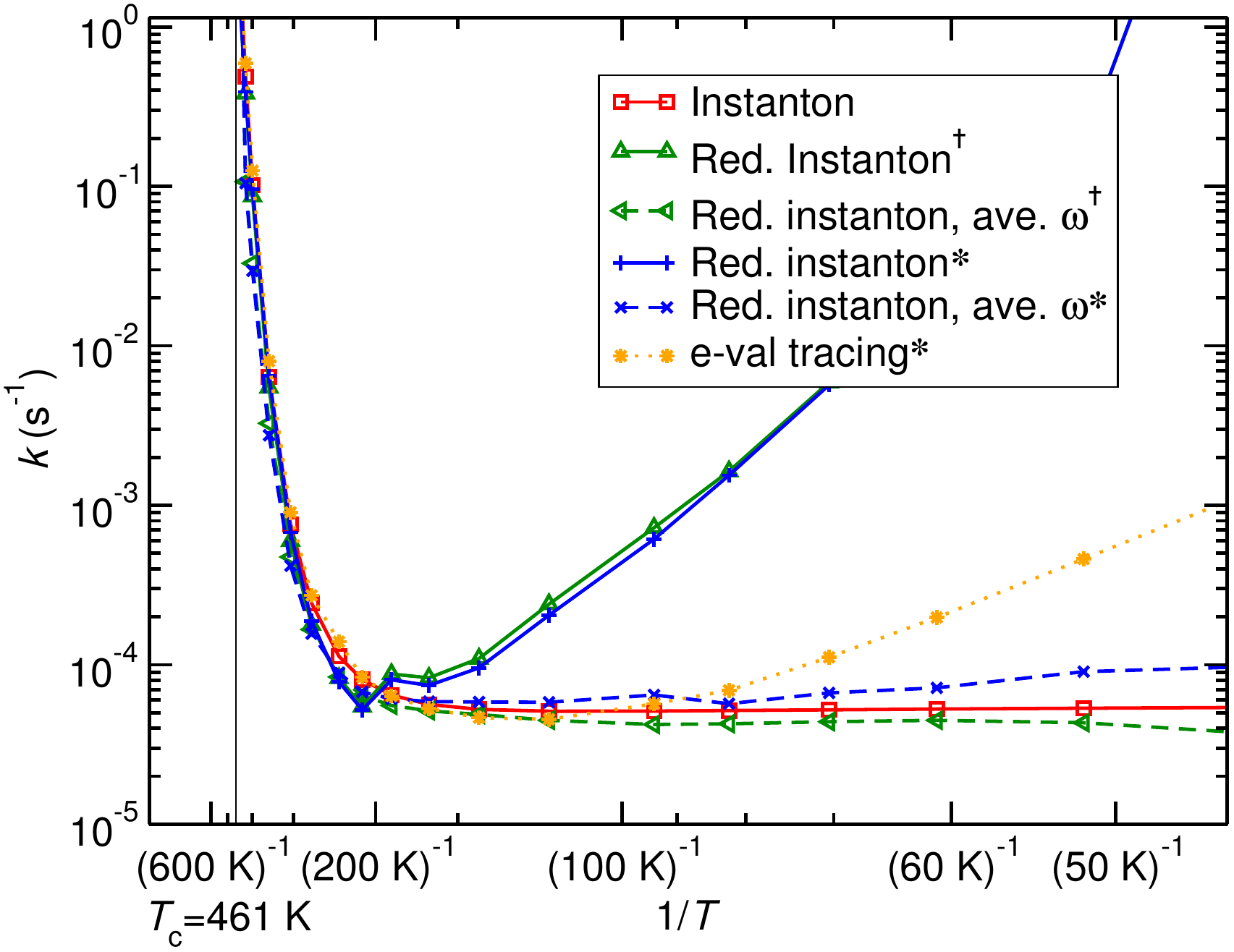}
	\includegraphics[width=8cm]{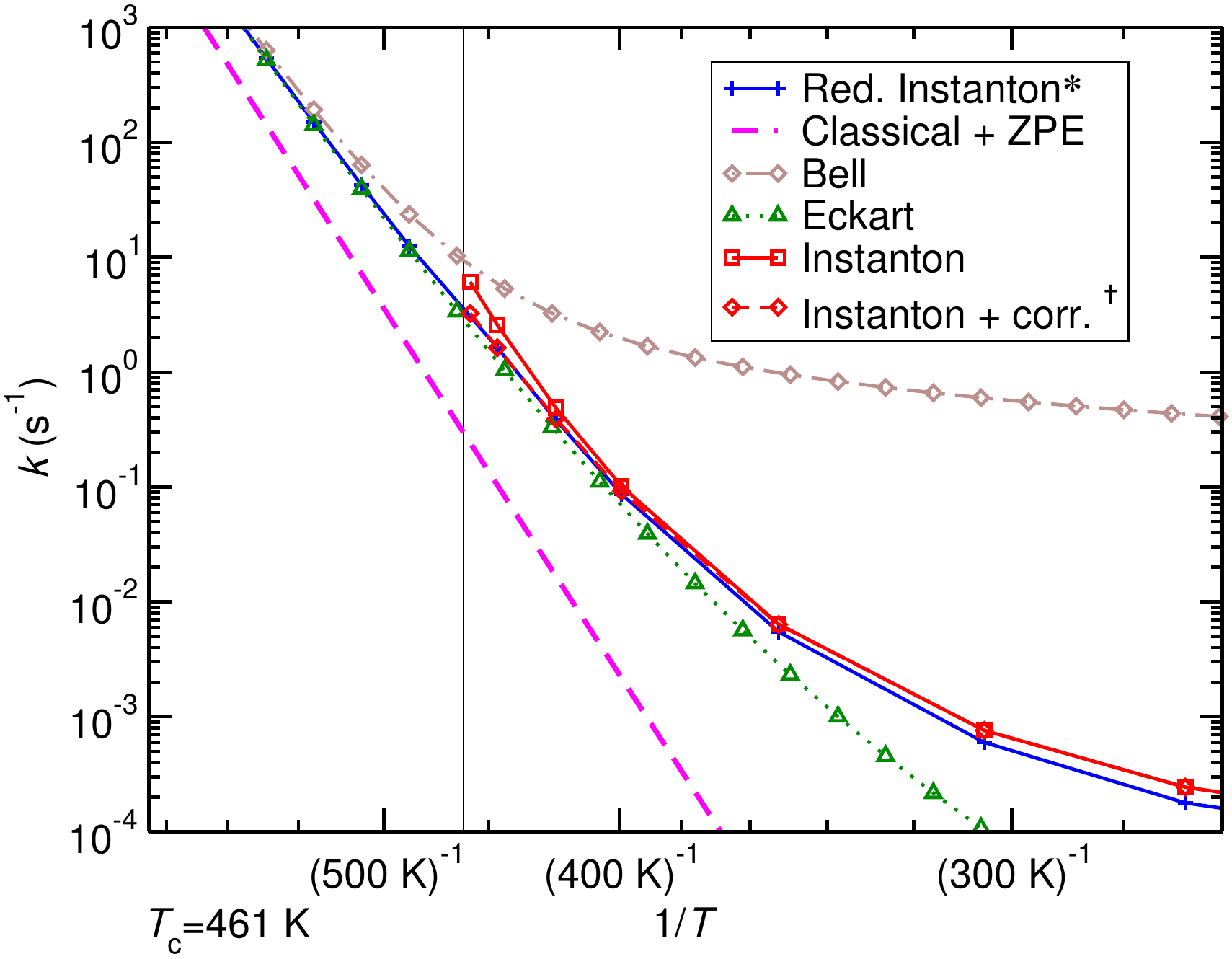}

      \caption{Rate constants 
      for the reaction of methylhydroxycarbene to acetaldehyde 
      below $T_\text{c}$: 
      Comparison between canonical instanton theory (red line) and 
      reduced instanton theory (blue \& green lines). The symbol $\ast$ 
      signifies $dE_b/d\beta$ was calculated using the finite difference
      method, $\dagger$ indicates $dE_b/d\beta$ was calculated using 
      equation \eqref{dEdB3}. Red diamonds 
    refer to \eqref{eq:inst_corr}.\label{fig:CH3COH_Tc}} 
    \end{figure}

    The most obviously striking feature in \figref{fig:CH3COH_Tc} 
    of the rate constant obtained 
    from reduced instanton theory for this reaction is the 
    failure of the stability matrix method below about 200~K. At this 
    temperature, the length of the instanton path has stretched to 
    a point that the deviation in the optimised instanton path to the 
    true instanton path is such that a naive solution to 
    \eqref{GYeq} would imply that a stable instanton orbit no longer 
    exists.
  
    What this means in concrete terms is there are no longer 
    two easily identifiable zero eigenvalues in the 
    spectrum of \eqref{sigma}, there remains however two 
    eigenvectors with a considerably larger projection on the instanton 
    path than all other eigenvectors. 
    We identify these eigenvectors and continue to remove their 
    eigenvalues from the sum in \eqref{sigma}. This is however 
    only a makeshift solution, as temperatures continue to decrease, 
    all the eigenvectors of the stability matrix exhibit a 
    non-negligible projection onto the instanton path, hence the 
    divergence of the green and blue lines.

    \setcounter{subfigure}{0}  
      
    Rate constants calculated using the frequency averaging approach 
    for $\sigma$ (left triangles \& crosses) are, as expected, 
    inaccurate when $T\lesssim T_\text{c}$. However as temperatures decrease 
    the rate constants become quite stable, and allow the reduced 
    instanton theory to produce valid results even at very low 
    temperatures. Above the crossover temperature in \figref{fig:CH3COH_Tc} 
    again a slight reduction in the rate constant
    is seen in comparison to canonical instanton theory near $T_\text{c}$.

  \subsection{\sffamily \large Bimolecular reaction: H$_2$ + OH $\to$ H + H$_2$O}

    The final component in our set of systems for method comparison 
    is the bimolecular reaction of   H$_2$ + OH $\to$ H + H$_2$O in the 
    gas phase. In the interest of providing a rigorous comparison 
    between methods, it is prudent to include a bimolecular type 
    reaction. The rate constants calculated here utilise the NN1 
    fitted potential energy surface.\cite{che13}
    
    This type of reaction is a prototypical reaction for four-atom 
    systems and is also of interest in astrochemistry, where, due to 
    low temperatures, reactions are dominated by tunnelling effects.
    \cite{Meisner2016}

    \begin{figure}
			\includegraphics[width=8cm]{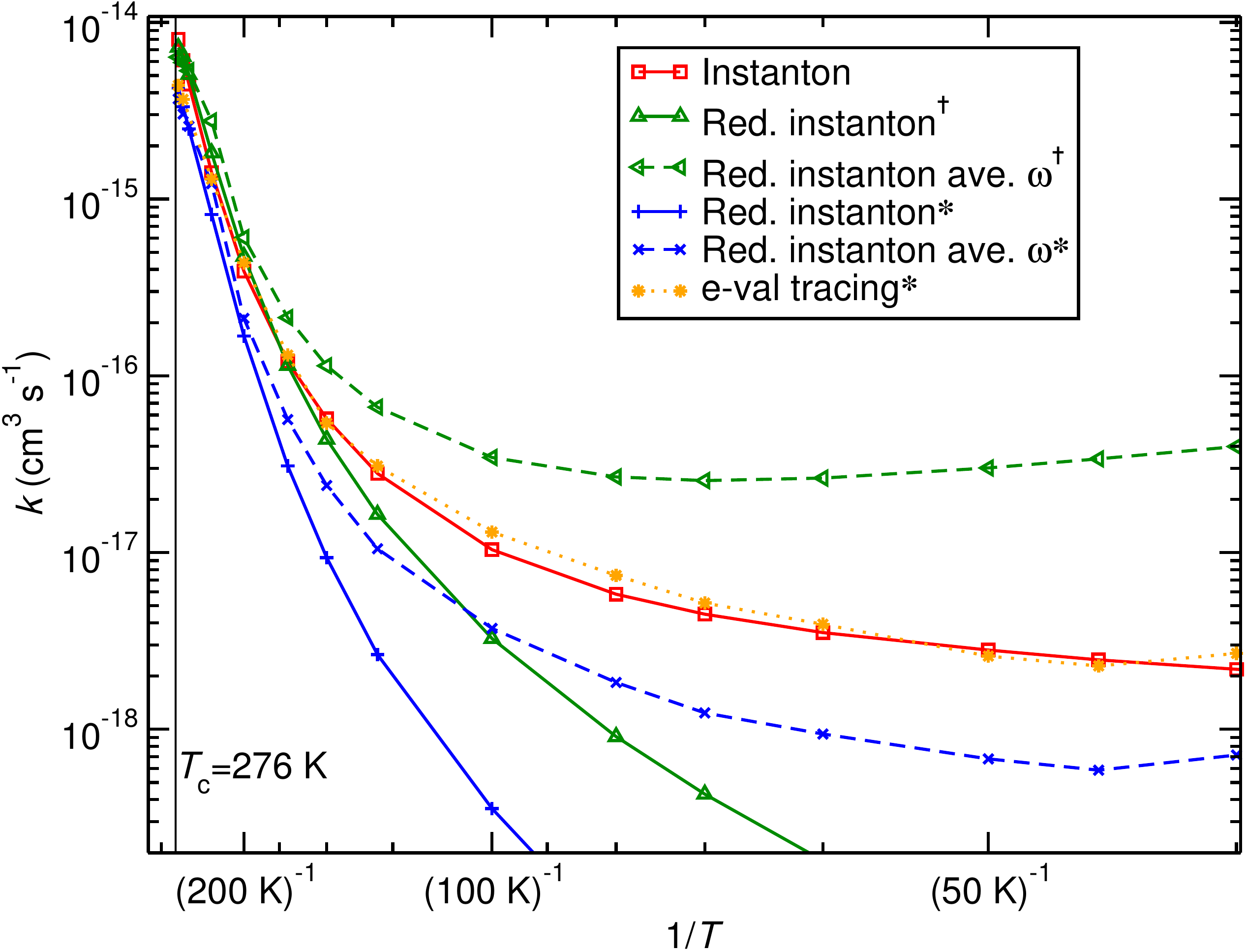}
			\includegraphics[width=8cm]{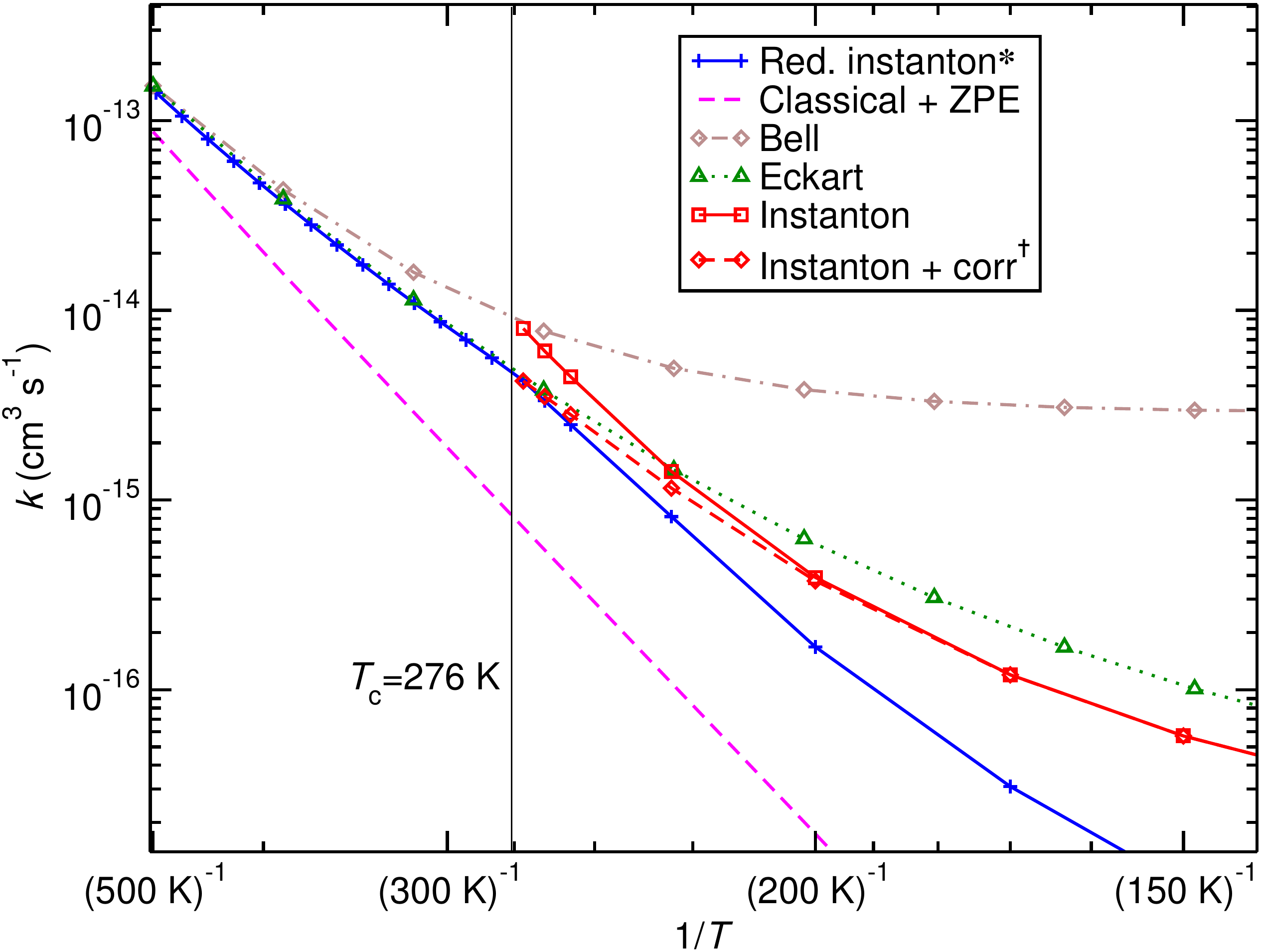}
      \caption{Reaction rate constant for the reaction of H$_2$ + OH $\to$ H +
        H$_2$O below $T_\text{c}$: Comparison between canonical instanton
        theory (red line) and reduced instanton theory (blue \& green
        lines). The symbol $\ast$ signifies $dE_b/d\beta$ was calculated using
        the finite difference method, $\dagger$ indicates $dE_b/d\beta$ was
        calculated using equation \eqref{dEdB3}. Red diamonds refer to
        \eqref{eq:inst_corr}.\label{fig:H2OH_Tc}}
    \end{figure}    

    As opposed to the other two test systems, the rate constants 
    obtained using the stability matrix approach tend to deviate  
    from the canonical instanton curve much faster. The rapid departure 
    of the blue curve in \figref{fig:H2OH_Tc} from the red curve 
    is evidence of this. However, since this system already has a quite 
    low crossover temperature ($T_\text{c}=276$ K), we 
    can expect the frequency averaging and eigenvalue tracing approaches
    to produce reasonable results for all temperatures below $T_\text{c}$. 
    As is visible in \figref{fig:H2OH_Tc}, usage of eigenvalue tracing 
    produces a smoother transition from below $T_\text{c}$ to above $T_\text{c}$ 
    than would be possible in reduced instanton theory without eigenvalue 
    tracing.
    At moderate to low temperatures, the rate 
    constants from reduced instanton theory without eigenvalue tracing in 
    \figref{fig:H2OH_Tc} (dashed, blue line) are consistently between approximately 
    one half and one order of magnitude smaller than those of 
    canonical instanton theory. 
    This is expected near the crossover 
    temperature where once more the rate constants are lower than 
    canonical instanton theory by about a factor of 2. At low 
    temperatures, both versions of the frequency averaged rate constants
    (left triangles \& crosses) deviate from canonical instanton theory (red curve)
    by around half an order of magnitude, however the rate constants 
    using the finite difference approach for $dE_b/d\beta$ (blue, dashed curve) 
    remain closer to those calculated using canonical instanton theory with decreasing
    temperatures.
    
    It is also worth noting that while in the other 
    test cases, eigenvalue tracing was not perceptibly better than 
    either of the other methods in calculating $\sigma$, here it 
    definitely outperforms both the rate constants obtained by 
    frequency averaging and those from solving the stability matrix 
    effectively at all temperatures. This is evidence that there are 
    still small, non-negligible frequencies which 
    invalidate \eqref{low_freq_cond}. Eventually 
    however, at lower temperatures, $\beta$ will increase to such a 
    point that \eqref{low_freq_cond} will be valid and the orange and 
    dashed, blue curve in the upper panel of \figref{fig:H2OH_Tc} 
    will meet.

		In each case studied, the rates calculated using 
		$k_\text{corr}$ (dashed, red lines) of \eqref{eq:inst_corr} 
		are very similar to those calculated using \eqref{Kryv_rate}, 
		particularly near the crossover temperature. Below the crossover 
		temperature, the discrepancies arise due to the methods used in 
		determining the stability parameters, as discussed in section 
		\ref{sec:stab_mat_de}.

\section{\sffamily \Large Discussion}\label{sec:disc}

  The reduced instanton theory permits the calculation of chemical 
  reaction rate constants to an accuracy which is comparable to the 
  established instanton theory at reduced computational expense.
  
  The method we have illustrated here involves a mixture of techniques 
  for the calculation of both the contribution to the action due to 
  tunnelling motion orthogonal to the instanton path and the on-the-fly 
  calculation of the rate of change of the tunnelling energy.   
  Regarding the action calculation, the stability matrix approach 
  produces acceptable results in a temperature range near but less than 
  the crossover temperature. In this temperature range, the tunnelling 
  path remains short and the projection matrices between eigenvectors of
  neighbouring Hessians is very close to the identity, allowing easy 
  identification of the zero stability parameters of \eqref{GYeq}. 
  In the frequency 
  averaging approach this idealised case of projections between 
  neighbouring Hessians is assumed fulfilled, cf. \eqref{ActionS_disc}, 
  and is responsible for its applicability as $T\to 0$~K. In the 
  stability matrix however, this condition must be fulfilled to within 
  a certain, system-dependent tolerance. As long as the projections 
  between eigenvector matrices of neighbouring Hessians is 
  approximately 1, this method can be used with confidence. The 
  frequency averaging approach is an inexpensive remedy when this 
  condition can no longer be fulfilled. Another such remedy would be to 
  rapidly increase the number of images comprising the instanton path, 
  yet this quickly renders the reduced instanton theory \emph{less} 
  computationally efficient than canonical instanton theory.

	In the range of temperatures well below the crossover temperature, 
	the average eigenvalue approach tends to remain stable and close to 
	the rates calculated using canonical instanton theory. In the reaction 
	H$_2$ + OH $\to$ H + H$_2$O however, the presence of low-frequencies shifts 
	the range where the average eigenvalue approach is fully applicable 
	to much lower temperatures than either of the two previous examples.  
  The eigenvalue tracing approach to 
  calculating $\sigma$, as outlined elsewhere\cite{McConnell2016} is 
  also useful as a cross-reference to check the validity of rate 
  constants calculated at low temperatures if the computational effort 
  of calculating rate constants from canonical instanton theory is 
  prohibitive.
  
\section{\sffamily \Large Conclusion}\label{sec:conc}
  There are two main advantages over canonical instanton theory in 
  using the reduced instanton theory. Firstly, for the 
  calculation of rate constants near and above $T_\text{c}$, the reduced 
  instanton theory is superior, providing a smooth transition from 
  temperatures below to above $T_\text{c}$ without the spurious kink found in 
  many Arrhenius plots created solely using canonical instanton theory. 
  Secondly, the calculation of rate constants in canonical 
  instanton theory requires the diagonalisation of a matrix of dimension
  $NP$. This diagonalisation can create a computational 
  bottleneck in calculating rate constants, potentially more so than 
  the calculation of the energies, gradients and Hessians at various 
  coordinates especially when a fitted potential energy surface is used.
  In the reduced instanton theory this bottleneck is avoided, there is 
  no need to diagonalise a matrix of dimension larger than $2N$. 
  
  An intermediate approach is to use canonical instanton theory
  with the correction factor of \eqref{eq:inst_corr}. It provides rate
  constants at the accuracy and reliability of canonical instanton theory
  while alleviating the overestimation of rate constants close to, but below
  $T_\text{c}$.

  We have also identified certain pitfalls in the theory, such as 
  the necessity of the frequency averaging approach for calculating 
  $\sigma$ as an insurance against the eventual failure of the 
  stability matrix when calculating rate constants at lower 
  temperatures.  There may be some utility in representing the instanton as a series 
  of periodic basis functions\cite{Kryvohuz2011} in order to continue locating the 
  zero eigenvalues in the stability matrix eigenvalue spectrum at 
  low temperatures, we are currently exploring implementing such a 
  solution in our next program version. 

  Also implemented is a simple method to approximate the rate of 
  change in the tunnelling energy with temperature. Since the formula 
  is analytic, there is no limit to its precision, though as shown best
  in the reaction H$_2$ + OH $\to$ H + H$_2$O precision is not a 
  guarantee of accuracy. Fortunately, any deviation in $dE_b/d\beta$ 
  is damped since it enters the rate equation under the square root. 
  Despite the seemingly large difference in $dE_b/d\beta$ calculated 
  by each method for H$_2$ + OH $\to$ H + H$_2$O 
  (cf. \figref{fig:dedb}) 
  the actual effect on rate constants, at temperatures ranging from 
  near $T_c$ to moderately low, is only about half an order of magnitude. 
  Clearly though at very low temperatures, this deviation increases, hence
  improving our simple approximation to 
  $dE_b/d\beta$ is one of the tasks for our group going forward.
    
  Finally, it is prudent to use canonical instanton theory in 
  conjunction with the reduced theory in order to benchmark rate 
  constants in the very low temperature range. It is 
  recommended that when calculating rate constants using the reduced 
  theory that, at the lowest temperature of interest, one should 
  perform a single calculation in the canonical instanton theory. 
  In doing so, one can visualise the results from using either the 
  stability matrix, or the frequency averaging approach. Though 
  it is certain that both approaches cannot calculate accurate 
  rate constants at arbitrarily low temperatures, the point at which 
  their results begin to strongly deviate from each other is entirely 
  system dependent. In this regard, one final calculation using 
  canonical instanton theory may help in deciding which of the 
  two methods is the more reliable in the temperature regime of 
  interest and should not add significantly to the 
  total computational expenditure.
 
\subsection*{\sffamily \large ACKNOWLEDGMENTS}
    This work was financially supported by the European Union's Horizon 
    2020 research and innovation programme (grant agreement No. 646717, 
    TUNNELCHEM).

 \bibliography{Q_Chem_Uni_St,jabref}

\end{document}